\documentclass{birkjour}
\usepackage{amsmath}
\usepackage{amssymb}
\usepackage{braket}
\usepackage{bbold}
\usepackage{pifont}
\usepackage{color}
\usepackage{dsfont}
\usepackage{hyperref}

\newcommand{\beqn}{\begin{eqnarray}}
\newcommand{\eeqn}{\end{eqnarray}}

\begin{document} 

\title{The Unified Standard Model}
\author{Brage Gording}
\address{Max-Planck-Institut f\"ur Physik (Werner-Heisenberg-Institut)\\
	F\"ohringer Ring 6, 80805 Munich, Germany}
\email{brageg@mpp.mpg.de}
\author{Angnis~Schmidt-May}
\address{Max-Planck-Institut f\"ur Physik (Werner-Heisenberg-Institut)\\
	F\"ohringer Ring 6, 80805 Munich, Germany}
\email{angnissm@mpp.mpg.de}

\begin{flushright}
	\hfill{MPP-2019-186} \vspace{20mm}
\end{flushright}

\begin{abstract}
	The aim of this work is to find a simple mathematical framework for our established description of particle physics.
	We demonstrate that the particular gauge structure, group representations and charge assignments of the
	Standard Model particles are all captured by the algebra M(8,\,$\mathds{C})$ of complex 8$\times$8 matrices. 
	This algebra is well motivated by its close relation to the normed division algebra of octonions.
	(Anti-)particle states are identified with basis elements 
	of the vector space M(8,\,$\mathds{C})$. Gauge transformations are simply described by the algebra acting on itself.
	Our result shows that all particles and gauge structures of the Standard Model 
	are contained in the tensor product of all four normed division algebras, 
	with the quaternions providing the Lorentz representations.
	Interestingly, the space M(8,\,$\mathds{C})$ contains two additional elements independent of 
	the Standard Model particles, hinting at a minimal amount of new physics.
\end{abstract}

\maketitle
\flushbottom



\section{Introduction}
\subsection{Motivation}

The Standard Model of Particle Physics is our well-established theory for elementary particles and their
fundamental interactions. It has been experimentally tested to a high precision and its last missing ingredient,
the Higgs boson, was detected in 2012~\cite{Aad:2012tfa, Chatrchyan:2012xdj}. 
While the establishment of this model has undoubtedly been one of the 
greatest achievements in fundamental science, the origin of its 
particular structure for particles and their interactions remains mysterious. 

The Standard Model is a quantum field theory invariant under the gauge group 
$G_\mathrm{SM}=\mathrm{SU}(3)\times \mathrm{SU}(2) \times \mathrm{U}_Y(1)$. 
In addition to the 12 gauge bosons, it contains 3 generations of fermions
and the complex Higgs scalar which all transform in certain representations of the gauge group.
More precisely, each generation of fermions appears in 1 (quark) triplet and 1 (lepton) singlet of SU(3). All fermions come in pairs where for left-handed fermions and the SU(3) singlet Higgs these pairs are in a doublet of SU(2). Additionally, fermions possess independent antiparticles.
In total, this adds up to $12 + 2\cdot2\cdot(3\cdot (3+1 ))+2 =62$ particle and anti-particle types in the Standard Model.

Except for the gauge bosons, nearly all particles carry $\mathrm{U}_Y(1)$ hypercharges whose values are (almost) fixed by 
gauge anomaly cancellation~\cite{Geng:1988pr, Minahan:1989vd, Babu:1989ex}.
Apart from this consistency requirement, the particular gauge structure, particle content and charge assignments 
do not possess a fundamental motivation and their origin remains unexplained within the Standard Model.
To many theoretical physicists such an ad-hoc choice made by Nature appears unsatisfactory
and has inspired the search for unifying structures that could be underlying our particle physics model
and explain (part of) its content.

Particularly popular examples are the so-called Grand Unified Theories (GUTs),
which are further motivated by gauge coupling unification at high energies~\cite{Georgi:1974sy, Weinberg:1974yx, Georgi:1974yf}. 
The original proposal by Georgi and Glashow dates back to 1974 and is based on SU(5) which contains the Standard Model gauge group $G_\mathrm{SM}$ as a subgroup~\cite{Georgi:1974sy}. This setup introduces a few additional fields but it cannot explain the origin of 3 generations of fermions. Moreover, unfortunately, the minimal realization of this neat proposal is ruled out by experimental limits on proton decay \cite{Miura:2016krn}. 

The gauge-mediated proton decay is avoided in the Pati-Salam model based on the gauge group SU(4) $\times$ SU(2) $\times$ SU(2)~\cite{Pati:1974yy}. This setup itself does not deliver a genuine unification but it can be embedded into the larger group Spin(10),\footnote{This GUT is often referred to as ``SO(10)'' by physicists. However, the Lie group used in the model is indeed Spin(10), the double cover of SO(10).}  which provides another well-studied GUT~\cite{Georgi:1974my}. A particularly nice feature of Spin(10) is that its 16 dimensional spinor representation can exactly incorporate one generation of the Standard Model fermions (including the unobserved right-handed neutrino). However, this comes at a price: spontaneously breaking the gauge symmetry down to $G_\mathrm{SM}$ requires a Higgs sector with representations of large dimensions. As a consequence, already the minimal model contains more than 100 new fields. This means that the number of particles is more than doubled with respect to the Standard Model. 
Moreover, requiring the presence of the Standard Model Higgs field again induces proton decay mediated by additional components in the corresponding Spin(10) representation~\cite{Georgi:1981vf, Grinstein:1982um, Masiero:1982fe}. Another deficiency is that, just like the 
Georgi-Glashow model, the Spin(10) model cannot explain why there are 3 generations of fermions.

Distinct from GUTs, noncommutative geometry is another approach at unification of the particle content of the Standard Model, \cite{Connes:1996gi,Chamseddine:2012sw,Chamseddine:1996zu,Chamseddine:2014uma}. Noncommutative geometry joins together gauge and spatial representations, with a finite dimensional space describing gauge representations joined to an infinite-dimensional Hilbert space. This approach has nice features such as being able to derive all bosonic particles as fluctuations of the inverse line element, with gravity being "external fluctuations" and the gauge and Higgs bosons being the "internal fluctuations". This approach also does not present a way in which to naturally obtain three generations of fermions. We refer the interested reader to the above-cited works.

This paper considers instead a shift of paradigm,
building on the ideas in recent works of 
C.~Furey~\cite{Furey:2010fm, Furey:2014iwa, Furey:2015tqa, Furey:2016ovx, Furey:2018yyy}. 
This programme extends earlier proposals 
in the literature~\cite{Gunaydin:1973rs, Gunaydin:1974fb, Gunaydin:1976vq, Barducci:1977cd, Casalbuoni:1979me, Casalbuoni:1979ta, DixonBook, Dixon:1988gn}
and its main idea is the following. Instead of embedding $G_\mathrm{SM}$ into a larger gauge group, like in GUTs,
we focus on very fundamental mathematical structures, called normed division algebras. 
Roughly speaking, such an algebra is characterized by
allowing for the operations of addition, subtraction, multiplication and division of its elements. 
There exist only 4 normed division algebras over the real numbers: the real numbers 
$\mathds{R}$ themselves, the complex numbers $\mathds{C}$, the quaternions $\mathds{H}$
and the octonions $\mathds{O}$. Furey has put forward arguments supporting the intriguing idea that this small set of division algebras
may contain the entire Standard Model, including its gauge structure, particle content, charge assignments,
and even its Lorentz representations.

A closer inspection of the division algebras and their interplay with particle physics is fundamentally 
motivated by the fact that $\mathds{R}$, $\mathds{C}$ and $\mathds{H}$ are already part of our 
established descriptions. The real and complex numbers appear essentially everywhere, 
while the quaternions are in fact realized in the form of Lorentz transformations.
It becomes an obvious question whether the fundamental description of Nature also 
requires the last division algebra, the octonions.

In this paper, we demonstrate that the structure of the complex octonions 
can describe the 62 independent particles and anti-particles of the Standard Model, 
including their gauge structure and charge assignments while incorporating minimal new degrees of freedom.
We emphasize that our main result is remarkably simple and understanding it requires no knowledge of division algebras. 
Any reader familiar with linear algebra and representation theory will be able to follow our work.

\subsection{Previously established results}
\label{previousresults}

More concretely, the general idea is to identify the structures of the Standard Model with the 
tensor product of all the normed division algebras,\footnote{Note that the real numbers are also
	taken into account since a tensor product with $\mathds{R}$ does not alter the structure of the algebra and thus
	$\mathds{D}= \mathds{R}\otimes\mathds{C}\otimes\mathds{H}\otimes\mathds{O}$.}
\begin{align}
\label{dixonalg}
\mathds{D}=\mathds{C}\otimes\mathds{H}\otimes\mathds{O},
\end{align}
referred to as the Dixon algebra \cite{Furey:2016ovx}. 
Following the program initiated by Furey, we will treat $\mathds{C}\otimes\mathds{H}$
and $\mathds{C}\otimes\mathds{O}$ separately. Specifically this means that we will consider 
some basis $\{\varepsilon_j\}$ of the complex vector space $\mathds{C}\otimes\mathds{H}$ and 
some basis $\{e_a\}$ of the complex vector space $\mathds{C}\otimes\mathds{O}$. 
The vector space $\mathds{D}$ will then be spanned by the basis elements 
$\{\varepsilon_j e_a\}$=$\{ e_a\varepsilon_j\}$ with complex-valued coefficients
or, equivalently, by $\{\varepsilon_j e_a,i\varepsilon_j e_a\}$ with real-valued coefficients.

It has already been shown that $\mathds{C}\otimes\mathds{H}$ contains 
all Lorentz representations present in the Standard Model~\cite{Furey:2010fm, Furey:2016ovx}.
Furey furthermore suggests that $\mathds{C}\otimes\mathds{O}$ will deliver the gauge structures, particle content, and charges. 
Therefore $\mathds{D}$ has the potential of providing a 
unified description of all Standard Model representations, of both Lorentz and gauge groups.

Let us briefly elaborate on this in a bit more detail. The algebra $\mathds{C}\otimes\mathds{H}$ 
contains the Lie algebra $\mathfrak{so}(1,3)$ of Lorentz transformations which generates the Lie group SO(1,3). 
In addition, $\mathds{C}\otimes\mathds{H}$ contains the Lorentz representations of covariant vectors, contravariant vectors, 
left- and right-handed Weyl spinors, field strength tensors, and scalars. 
An additional feature is that, in this algebra, left- and right-handed Weyl spinors are related simply by complex conjugation, 
which is not the case when these objects are represented in spinor space \cite{Furey:2016ovx}. 
Since the anti-particle of a right-handed particle is left-handed, 
complex conjugation can here be interpreted as interchanging particle and anti-particle states.\footnote{In fact, 
	complex conjugation will also flip the spin, such that the complex conjugate of a right-handed 
	Weyl spinor with spin up is a left-handed Weyl spinor with spin down.}

Turning our attention to $\mathds{C}\otimes\mathds{O}$, we first note that this algebra is non-associative. However, part of the octonions' structure can be captured in an associative algebra, 
as described in detail in \cite{Furey:2016ovx, DixonBook}. 
Explicitly, non-associativity implies that if we take three elements 
$a,b,c \in\mathds{C}\otimes\mathds{O}$ the product $abc$ is ill defined, as generically $(ab)c\neq a(bc)$.
This ambiguity in the expression $abc$ can be removed by demanding multiplication to always happen from right to left,
\begin{align}
abc\,f\equiv a(b(c(f))),\quad\forall f\in\mathds{O}\,.
\end{align}
This definition renders the expression $abc$ well-defined and corresponds to the composition of three maps 
$a,b,c:\mathds{C}\otimes\mathds{O}\to\mathds{C}\otimes\mathds{O}$. 
The vector space of all such maps, together with the operation of map composition, forms an associative algebra, 
denoted $\mathds{C}\otimes\overleftarrow{\mathds{O}}$. 
Since there is good reason to believe that Standard Model physics is described by associative algebras, 
Furey focusses on the algebra $\mathds{C}\otimes\overleftarrow{\mathds{O}}$, 
which captures the multiplicative structure of the octonions.\footnote{The associative algebra $\mathds{H}$, 
	is isomorphic to the algebra of maps $\overleftarrow{\mathds{H}}\cong\mathds{H}$.
	Both $\overleftarrow{\mathds{O}}$ and $\mathds{H}$ are thus subalgebras of 
	$\overleftarrow{\mathds{D}}\cong\mathds{C}\otimes\mathds{H}\otimes\overleftarrow{\mathds{O}}$.
	Moreover, $\overleftarrow{\mathds{O}}$ 
	is equivalent to the algebra of adjoint actions on the octonions \cite{DixonBook}.
	It follows that $\overleftarrow{\mathds{D}}$ is the algebra of adjoint actions on $\mathds{D}$.}

It can be shown that the algebra $\mathds{C}\otimes\overleftarrow{\mathds{O}}$
is isomorphic to the complex Clifford algebra $\mathds{C}l(6)$~\cite{Furey:2018yyy,DixonBook}.
In this Clifford algebra, Furey has successfully identified some of the Standard Model gauge structures
using spinor construction. 
In Ref.~\cite{Furey:2015tqa,Furey:2018yyy}, one full generation of leptons and quarks along with their anti-particles are found.
They are shown to transform appropriately under the SU(3)$\times\text{U}_{\mathrm{em}}$(1) gauge groups which are also 
generated by elements of the algebra. 
Weak-isospin states are identified but their full SU(2) transformations are missing. 
In Ref.~\cite{Furey:2014iwa, Furey:2018yyy}, Furey finds 3 generations of Standard Model fermions 
and their anti-particles in correct SU(3) representations. 
However, this identification requires two separate $\mathfrak{su}(3)$ Lie algebras. There is no SU(2) gauge group
and no Higgs particle. Hence, a full identification of the Standard Model gauge structures and representations
inside $\mathds{C}\otimes\overleftarrow{\mathds{O}}$ is pending.

A remarkable result of Furey's analysis is that, in the algebra $\mathds{C}\otimes\overleftarrow{\mathds{O}}$,
the electric charges emerge as eigenvalues of a number operator~\cite{Furey:2015tqa}.
This offers a straightforward explanation for the quantization of electric charge.

It is worth noting that similar ideas have been explored in other literature. 
The structure for quarks was first related to the split octonions in Ref.~\cite{Gunaydin:1973rs, Gunaydin:1974fb}. 
Ref.~\cite{Stoica:2017iuo} found one generation of leptons and quarks, but without SU(2) symmetry. 
In a different approach, Ref.~\cite{Gillard:2019ygk} 
studied a larger algebra which contains 3 copies of the octonions, 
one for each generation.
For more work on the relation of octonions and particle physics, see \cite{Baez:2001dm, Baez:2009xt, Manogue:2009gf, Boyle:2014wba, Anastasiou:2014zfa,Todorov:2018yvi,Todorov:2019hlc}.
The octonion algebra has also been studied in the context of M-Theory compactifications~\cite{Gunaydin:1983mi, deWit:1983gs}.

\section{Main Result}\label{mainresults}

In all that follows, we will invoke a further isomorphism which is made explicit in appendix \ref{appendix:matrixisomorphism}.
It allows us to make the identification 
$\mathds{C}\otimes\overleftarrow{\mathds{O}}\cong\mathds{C}l(6)\cong \text{M}(8,\mathds{C})$.
In other words, we will simply work with the familiar algebra $\text{M}(8,\mathds{C})$ of 8$\times$8 complex matrices
and their matrix multiplication.
Therefore the reader only needs to know basic linear algebra (and some basic representation theory) 
to understand the results of this paper.

\subsection{Overview}

$\text{M}(8,\mathds{C})$ is a complex vector space of 64 dimensions. 
As such, it can be spanned by 64 independent basis elements. 
As mentioned already in the introduction, ignoring Lorentz structures like chirality, 
the Standard Model contains 62 independent particle types. That is, one particle for each dimension of our vector space, with only two new particle components. 

We find that the vector space $\text{M}(8,\mathds{C})$ 
decomposes into a set of linearly independent subspaces which can be assigned to the different 
particles of the Standard Model. Specifically,%
\footnote{Note that there is an overall factor of $\mathds{C}$ in this decomposition. By (\ref{dixonalg}) the same complex field enters into the Lorentz representations discussed in Appendix \ref{lorentztransfweyl}. However, we are here only focusing on a subalgebra of this tensor product, namely $\mathds{C}\otimes\mathds{O}$. As such, we cannot currently comment on any physical significance of the overall factor of $\mathds{C}$ in the decomposition. It is known, \cite{Furey:2010fm}, that for vector fields such an overall $\mathds{C}$ corresponds to the combination of hermitian and anti-hermitian vectors in $\mathds{C}\otimes\mathds{H}$.}
\begin{align}
\label{decompM8}
\text{M}(8,\mathds{C})= \mathds{C}\,\otimes\,\Big[\mathfrak{su}(3)\,\oplus\,\mathfrak{su}(2)\,\oplus\,\mathfrak{u}_Y(1)
\,\oplus\, 3\cdot\big(F_3\,\oplus \,F_{\bar{3}}\,\oplus\, F_1\,\oplus \,F_{\bar{1}}\big)
\,\oplus \,F_\phi\,\oplus P_\mathrm{BSM}\Big]\,.
\end{align}
The first three subspaces correspond to the gauge fields 
which satisfy the commutation relations dictated by their respective Lie algebras. 
The $F_i$ subspaces are shown to all transform as doublets under the group generated by $\mathfrak{su}(2)$. 
Under the group generated by $\mathfrak{su}(3)$,
$F_3$ transforms as a triplet, $F_{\bar{3}}$ as an anti-triplet, and 
$F_1$, $F_{\bar{1}}$, $F_\phi$ as singlets. 
Carrying the correct $\mathfrak{u}_Y(1)$ hypercharge, we identify the 
$F_3$'s as quark colour triplets, $F_{\bar{3}}$'s as quark anti-colour triplets, 
$F_1$'s as leptons, $F_{\bar{1}}$'s as anti-leptons, and $F_\phi$ as the Higgs doublet. 
The remaining space $P_\mathrm{BSM}$ is a two-dimensional complex subspace of M(8,$\mathds{C}$), 
linearly independent from the Standard Model particles, and thus a candidate for new physics.

The algebra $\text{M}(8,\mathds{C})$ is extraordinarily simple and powerful.
Each particle is identified with a basis vector and all gauge group representations 
are linearly independent subspaces of the algebra.
The gauge transformations of particle representations simply correspond to the algebra acting on itself.
That is, all particles, gauge structures, and charge assignments of the Standard Model can be described 
purely in terms of a 64-dimensional matrix algebra. 

We emphasise that $\text{M}(8,\mathds{C})$ is the smallest matrix space which can incorporate the different particles of the Standard Model. 
This has some interesting implications.
For instance, the Standard Model with its gauge group and particle content 
can be identified with $\text{M}(8,\mathds{C})$ 
if and only if it contains no more than 3 generations of fermions.

While our approach is similar in spirit to GUTs, in that we seek a single structure to explain the Standard Model, 
it has several advantages:
\begin{enumerate}
	\item Our setup introduces a minimal amount of extra degrees of freedom, namely two new particle types.
	
	\item The number of (anti-)particles is identified with the dimension of the algebra. In this sense, 
	the algebra can explain the absence of a 4th generation of fermions.
	
	\item Gauge generators and particle representations are all elements of the same algebra;
	gauge transformations of all fields are described by the algebra acting on itself. Thus, 
	at the level of $\text{M}(8,\mathds{C})$, there is no 
	fundamental distinction between gauge bosons, fermions and the Higgs boson. Their behaviour 
	under gauge transformations is largely encoded in the structure of the algebra.

	\item The algebra $\text{M}(8,\mathds{C})$ is naturally part of a larger algebra, 
	$\overleftarrow{\mathds{D}}$, which also includes the Lorentz structures of the Standard Model.
	Moreover, $\mathds{D}$ exhausts the set of all 4 normed division algebras.

\end{enumerate}
There are additional appealing features of the algebra, like the simple relationship between particles and anti-particles, which we will elaborate on later.

\subsection{Technical summary}

In the following we discuss in more detail how the space $\text{M}(8,\mathds{C})$ decomposes, 
as per (\ref{decompM8}), into the particle content and gauge structure of the Standard Model.

\subsubsection{Choosing a basis of $\text{M}(8,\mathds{C})$}
\label{basis}

Let $R_I$ with $I=1,\hdots,8$ be a complete set of basis vectors of $\mathds{C}^8$, chosen such 
that their inner product is\footnote{These basis vectors also satisfy 
${R_I}^{\bar{*}}\equiv \begin{pmatrix}  
	0 &\eta
	\\
	\eta &0
\end{pmatrix} {R_I}^* =R_{I+4}$, where $\eta=\mathrm{diag}(1,-1,-1,-1)$.
 The $\bar{*}$ is an operation in M(8,$\mathds{C}$) that corresponds to complex conjugation in $\mathds{C}\otimes\overleftarrow{\mathds{O}}$, as described in appendix \ref{appendix:matrixisomorphism}.}
\beqn
\label{orthoR}
{R_I}^\dagger R_J=\delta_{IJ}\,.
\eeqn
We furthermore define 8 vectors $\{V^+_a,V^-_a\}$ with $a=0,\hdots, 3$ and
express them as linear combinations of the above basis vectors, 
\beqn\label{vlinc}
V^\pm_a= \sum_{I=1}^8 a^\pm_{aI} R_I\,.
\eeqn
The complex coefficients $a_{aI}^\pm$ are chosen such that the $V^\pm_a$ are linearly independent 
and their inner products satisfy\footnote{Additionally, 
we will require that $(V^\pm_a)^{\bar{*}}=V^\mp_a$
which implies $\left(a_{aI}^+\right)^* = a_{a(I+4)}^-$. }
\beqn
\label{orthoV}
(V^\pm_a)^\dagger V^\pm_b=\delta_{ab}\,,\qquad
(V^\pm_a)^\dagger V^\mp_b=0\,.
\eeqn
Note that the vectors $\{V^+_a,V^-_a\}$ form another orthonormal basis of $\mathds{C}^8$.

Out of the basis vectors $R_I$ one can construct a basis $M_{IJ}$ of $\text{M}(8,\mathds{C})$ 
using the outer product, 
\beqn\label{genbas}
M_{IJ}=R_I R_J^\dagger\,, \qquad I,J=1,\hdots,8\,.
\eeqn
The independent particle types of the Standard Model will be identified with 62 linearly independent combinations 
of these basis elements.

\subsubsection{Identification with particles}
\label{summaryresults}
Using the basis vectors defined in the previous subsection, we identify the Standard Model content 
of the algebra $\text{M}(8,\mathds{C})$ as follows.

\paragraph{The SU(3) generators.}
The $\mathfrak{su}(3)$ Lie algebra is spanned by the generators,
\begin{align}
\label{su3gen}
\nonumber
&\lambda_1 = R_2\left(R_1\right)^\dagger + R_1\left(R_2\right)^\dagger - R_6\left(R_5\right)^\dagger - R_5\left(R_6\right)^\dagger\,,
\\\nonumber
&\lambda_2 = iR_2\left(R_1\right)^\dagger - i R_1\left(R_2\right)^\dagger +i R_6\left(R_5\right)^\dagger - i R_5\left(R_6\right)^\dagger\,,
\\\nonumber
&\lambda_3 = R_1\left(R_1\right)^\dagger - R_2\left(R_2\right)^\dagger - R_5\left(R_5\right)^\dagger + R_6\left(R_6\right)^\dagger\,,
\\\nonumber
&\lambda_4 = R_1\left(R_3\right)^\dagger + R_3\left(R_1\right)^\dagger - R_5\left(R_7\right)^\dagger - R_7\left(R_5\right)^\dagger\,,
\\\nonumber
&\lambda_5 = iR_3\left(R_1\right)^\dagger - iR_1\left(R_3\right)^\dagger + iR_7\left(R_5\right)^\dagger - iR_5\left(R_7\right)^\dagger\,,
\\\nonumber
&\lambda_6 = R_3\left(R_2\right)^\dagger + R_2\left(R_3\right)^\dagger - R_7\left(R_6\right)^\dagger - R_6\left(R_7\right)^\dagger\,,
\\\nonumber
&\lambda_7 = iR_3\left(R_2\right)^\dagger - iR_2\left(R_3\right)^\dagger + iR_7\left(R_6\right)^\dagger - iR_6\left(R_7\right)^\dagger\,,
\\
&\lambda_8 = \frac{1}{\sqrt{3}}\left[R_1\left(R_1\right)^\dagger + R_2\left(R_2\right)^\dagger - 2 R_3\left(R_3\right)^\dagger - R_5\left(R_5\right)^\dagger -R_6\left(R_6\right)^\dagger +2 R_7\left(R_7\right)^\dagger\right]\,.
\end{align}
Using the orthonormality relations (\ref{orthoR}), it is easy to verify that these elements indeed satisfy 
the $\mathfrak{su}(3)$ commutation relations.

\paragraph{The U(1) generator.}
The hypercharge generator is given by,
\beqn
\label{hypgen}
Y = R_8\left(R_8\right)^\dagger-R_4\left(R_4\right)^\dagger + 
\frac13\sum_{I=1}^3 R_{I}\left(R_I\right)^\dagger-\frac13\sum_{I=5}^7 R_{I}\left(R_I\right)^\dagger\,.
\eeqn

\paragraph{The SU(2) generators.}
The $\mathfrak{su}(2)$ Lie algebra is spanned by the generators,
\begin{align}\label{su2genparticle}
&T_1 = \sum_{a=0}^3  V_a^-\left(V_a^+\right)^\dagger + V_a^+\left(V_a^-\right)^\dagger \,,
\nonumber
\\\nonumber
&T_2 = \sum_{a=0}^3 i V_a^+\left(V_a^-\right)^\dagger -i V_a^-\left(V_a^+\right)^\dagger\,,
\\
&T_3 = \sum_{a=0}^3  V_a^+\left(V_a^+\right)^\dagger - V_a^-\left(V_a^-\right)^\dagger.
\end{align}
As for $\mathfrak{su}(3)$, the orthonormality relations (\ref{orthoV}) imply the
desired $\mathfrak{su}(2)$ commutation relations.

\paragraph{The fermions.}
The 16 elements of $\text{M}(8,\mathds{C})$ which transform as one generation of 
particles and antiparticles under the gauge groups correspond to
\beqn
\label{sum:gen}
\Big\{R_I\left(V^\pm_a\right)^\dagger\Big| I=1,...,8\Big\}\,,
\eeqn
where the index $a\in\{1,2,3\}$ labels the generation.
The elements with $I\in\{1,2,3,4\}$ 
give the particle states while those with $I\in\{5,6,7,8\}$ give the antiparticle states. 
 
\paragraph{The Higgs.} 
The 2 elements describing the Higgs doublet are
\beqn
\label{sum:higgs}
\Big\{R_4\left(V_\phi^\pm\right)^\dagger\Big\},
\eeqn
for a linear combination,
\beqn
\label{Vphidef}
V_\phi^\pm=\sum_{a=0}^3 h_a^\pm V^\pm_a\,,
\eeqn
where $h_a^\pm\in \mathds{C}$, and $\left(h_a^+\right)^*=h_a^-$ with $h_0^\pm\neq 0$.

\paragraph{The transformation laws.}
We assign the SU(3) transformation property
\beqn
\label{su3transf}
K\to e^{i\lambda_I} K\,,
\eeqn
for any $\lambda_I$ in (\ref{su3gen}) and for all elements $K$ in (\ref{sum:gen}) and (\ref{sum:higgs}).
Using (\ref{orthoR}) it is easy to verify that
the elements in (\ref{sum:gen}) with index $I\in\{1,2,3\}$ 
transform as a SU(3) triplet, those with $I\in\{5,6,7\}$ as an anti-triplet, and 
those with $I=4,8$ as singlets. The elements in (\ref{sum:higgs}) transform as singlets as well.

For U$_Y$(1) we demand a similar transformation law,
\beqn
\label{weaktransforms}
K\to e^{iY} K\,,
\eeqn
for all elements $K$ in (\ref{sum:gen}) and (\ref{sum:higgs}).
We see that elements with $I=1,2,3$ have charge $\frac13$, those with $I=5,6,7$ have charge $-\frac13$, 
the element with $I=4$ has charge $-1$, and that with $I=8$ has charge $1$. 
It can moreover be shown that the hypercharge generator (\ref{hypgen}) 
commutes with all SU(3) generators in (\ref{su3gen}).

Turning to SU(2), all elements $K$ of (\ref{sum:gen}) and (\ref{sum:higgs}) 
carry weak isospin charge, denoted by the superscript $\pm$. 
We demand that they transform as follows under the SU(2) generated by (\ref{su2genparticle})
with $k=1,2,3$,
\beqn
\label{su2transfpart}
K\to Ke^{iT_k}
\eeqn
for $I\in\{1,2,3,4\}$, and
\beqn
\label{su2transfantipart}
K\to K^{-iT_k^{\bar{*}}}
\eeqn
for $I\in\{5,6,7,8\}$.
At this point it may seem strange that our SU(2) transformation is acting from the right, 
unlike the SU(3) and $\text{U}_Y(1)$ transformations.\footnote{In fact, 
in the standard formulation of particle physics, such an action from the right 
is not even well defined: Particles in the fundamental representation of the gauge group  
are represented by column vectors and can only be acted upon by matrices from the left. 
Here, the situation is different
since all our particles correspond to elements of the matrix algebra. 
Hence they can be multiplied by matrices both from the left and from the right.}
We will show later that not only is this natural, 
it is required in order for the elements in (\ref{sum:gen}) and (\ref{sum:higgs}) to have the correct transformation properties.
As discussed in detail in appendix \ref{app:bogg}, this also implies that the action of SU(2) commutes with those
of SU(3) and U(1), even though the generators (\ref{su2genparticle}) do not commute with (\ref{su3gen})
and (\ref{hypgen}).

\paragraph{Linear independence.}

The necessary and sufficient conditions for linear independence of (\ref{su3gen})-(\ref{sum:higgs}) are
\begin{align}
\label{independencedonditions}
\begin{pmatrix}
a_{01}^+&a_{02}^+&a_{03}^+
\end{pmatrix}^\mathrm{T}~&\not\propto~
\begin{pmatrix}
(a_{05}^+)^*&(a_{06}^+)^*&(a_{07}^+)^*
\end{pmatrix}^\mathrm{T}\,,
\nonumber\\
\begin{pmatrix}
a_{01}^+&a_{02}^+&a_{03}^+
\end{pmatrix}^\mathrm{T}~&\not\perp~
\begin{pmatrix}
(a_{05}^+)^*&(a_{06}^+)^*&(a_{07}^+)^*
\end{pmatrix}^\mathrm{T}\,,
\qquad
h^\pm_0\neq0\,.
\end{align}
Neither of the two $\mathds{C}^3$ vectors above may be the zero vector.

In addition we must pick the $a_{aI}^+$, for $a\in\{1,2,3\}$, such that the orthormality relations in 
(\ref{orthoV}) are satisfied. 
The conditions (\ref{independencedonditions}), which are derived in section \ref{appendix:linearindependence}, 
are very weak and still allow for a lot of freedom in choosing our basis elements. In fact, the set of $a_{aI}^\pm$ 
still contains 28 real parameters after all conditions for orthogonality, conjugation and linear independence 
have been imposed.

The linear independence of (\ref{su3gen})--(\ref{sum:higgs}) provides the decomposition of the algebra in (\ref{decompM8}). The two remaining linearly independent elements spanning $P_\mathrm{BSM}$ lying outside the Standard Model particle content.

\section{Derivation}

We will now show how to arrive at the linearly independent elements identified
with the Standard Model particles in the previous section.

Let us begin by stating our basic assumptions. 
The matrices in M(8,$\mathds{C}$) are just mathematical objects which a priori carry no physical meaning.
It is obvious that we need to invoke additional assumptions in order to make a connection to particle physics. 
Thus, the following demands should be interpreted as the way in which physical meaning 
is assigned to the algebra M(8,$\mathds{C}$).

\begin{itemize}

	 \item[1.)] 26 elements of M(8,$\mathds{C}$) will be assigned to the fermions and the Higgs doublet.
	 Each of these particle elements of the algebra has an associated anti-particle element with opposite charge assignments. 
	 The two are related by complex conjugation, but do not necessarily describe independent particles.\footnote{The complex conjugation here refers to
	 complex conjugation in the Clifford algebra and hence to the
	 $\bar{*}$ operator on M(8,$\mathds{C}$). This is discussed in detail in appendix \ref{appendix:matrixisomorphism}.}
	 
	\item[2.)] Gauge transformations are described by letting the algebra act on itself. 
	For each factor in the Standard Model gauge group, we impose transformation properties 
	which are the same for all (anti-)particle states.
	 
\end{itemize}
We will start with the first of these requirements and (more or less arbitrarily) assign a set of 26 
linearly independent elements of M(8,$\mathds{C}$) to all weak isospin doublets of the Standard Model. 
In the second step, we impose universal gauge transformation laws for all our (anti-)particle states
and thereby identify the gauge generators.

\subsection{Fermions and Higgs}
\label{sec:fundrep}

Starting from the basis vectors introduced in section~\ref{basis},
consider the following elements of M(8,$\mathds{C}$),
\beqn
\Big\{R_I\left(V^+_a\right)^\dagger\Big| I=1,...,4\Big\}\,.
\eeqn
For $a=1,2,3$, these will be identified with the fermions of the Standard Model.
As will become clear later, the ``$+$" index denotes that the particles corresponding to these elements
have weak isospin value ``up". Their companion particles 
of weak isospin ``down" are denoted by
\beqn
\Big\{R_I\left(V^-_a\right)^\dagger\Big| I=1,...,4\Big\}\,.
\eeqn
Hence the set of elements
\beqn
\label{particlegen}
\Big\{R_I\left(V_a^\pm\right)^\dagger\Big| I=1,...,4\Big\}
\eeqn
describes one generation of weak isospin doublets, 
where the generation is labelled by $a\in\{1,2,3\}$.

Next we will assign SU(3) and U$_Y$(1) charges to these basis elements. 
We choose, arbitrarily, to assign the SU(3) charges (red, green, blue) to $I=(1,2,3)$ respectively, 
and to make $I=4$ a singlet of SU(3). For hypercharges consistent with the Standard Model charge allocations we must then assign the 
indices $I=(1,2,3)$ a hypercharge of $\frac13$ and the index $I=4$ a hypercharge of $-1$.

Having described three generations of particles, we now identify their respective anti-particles. 
Since we want particles and antiparticles to have opposite electroweak charge
and be related via the complex conjugation operation $\bar{*}$,
we choose the vectors $V_a^\pm$ such that,
\beqn
\label{conjugateV}
\left(V_a^\pm\right)^{\bar{*}}
=V_a^\mp\,.
\eeqn
We then have that,
\beqn
\label{CCparticlegen}
\left(R_I\left(V_a^\pm\right)^\dagger\right)^{\bar{*}} = \left(R_I\right)^{\bar{*}}\left(V_a^\mp\right)^\dagger\,.
\eeqn
Since particle and antiparticle states have different charges they must be described by linearly independent basis vectors,  
(\ref{CCparticlegen}) must be linearly independent from (\ref{particlegen}). 
We thus demand our basis vectors to satisfy,
\beqn
\label{conjugateR}
\left(R_I\right)^{\bar{*}}=R_{I+4}.
\eeqn
Then the antiparticle states corresponding to (\ref{particlegen}) are
\beqn
\label{antiparticlegen}
\Big\{R_I\left(V_a^\mp\right)^\dagger\Big| I=5,...,8\Big\}\,.
\eeqn
To summarize, each $a\in\{1,2,3\}$ gives one full generation of particles and antiparticles of the form, 
\beqn\label{sum:gen2}
\Big\{R_I\left(V^\pm_a\right)^\dagger\Big| I=1,...,8\Big\}\,.
\eeqn
Here, $I=8$ denotes a singlet of SU(3) with hypercharge $1$, and $I=(5,6,7)$ all have hypercharge $-\frac13$ with SU(3) charges (anti-red, anti-green, anti-blue) respectively.

Next we turn to the Higgs doublet which must have the same charge assignment as the SU(3) singlet of (\ref{particlegen}). 
This leads us to elements of the type
\beqn
\label{particlehiggs}
\Big\{R_4\left(V_\phi^\pm\right)^\dagger\Big\}\,.
\eeqn
Here $V_\phi^\pm$ are linear combinations of the $V^\pm_a$ which must include $V_0^\pm$ for (\ref{particlehiggs}) 
to be linearly independent from (\ref{sum:gen2}).
The complex conjugate of (\ref{particlehiggs}) yields another pair of elements,
\beqn
\label{antiparticlehiggs}
\Big\{R_8\left(V_\phi^\mp\right)^\dagger\Big\}\,.
\eeqn
While these have a charge assignment opposite to the Higgs particle, 
they cannot be made linearly independent from the rest of the Standard Model particle content.
Hence the elements in (\ref{antiparticlehiggs}) cannot describe an independent particle. 
This means that our Higgs doublet does not have an independent antiparticle state and 
only corresponds to the basis elements (\ref{particlehiggs}).
This finding matches the Standard Model: while the Yukawa interactions require both the Higgs doublet and its conjugate, the same two complex parameters which describe the Higgs appear in the conjugate doublet (see, for instance, Ref.~\cite{Peskin:1995ev}). Therefore, in the Standard Model the conjugate doublet does not provide an independent particle, by construction. Conversely, in the algebra M(8,$\mathds{C}$) the lack of an independent conjugate doublet is derivable, following directly from the identification of particles as linearly independent basis elements.

\subsection{The gauge generators}

In this section we will impose the desired gauge transformation properties of the particle states, 
(\ref{particlegen}) and (\ref{particlehiggs}). 
Before we begin, let us consider an arbitrary gauge transformation, given by some operator $\mathcal{O}$.
We require this transformation to commute with the complex conjugation ${\bar{*}}$ which
exchanges particle and anti-particles. In other words, we would like particles and anti-particles 
to obey the same transformation law.
Then for any $K$ in (\ref{particlegen}) or (\ref{particlehiggs}), we must have that
\beqn
\label{conjugategaugetransform}
\left(\mathcal{O}K\right)^{\bar{*}} \overset{!}{=} \mathcal{O}K^{\bar{*}}.
\eeqn
In the following this will be used to restrict the form of our gauge generators.

\subsubsection{SU(3) transformations}

To derive the $\mathfrak{su}(3)$ generators of transformations, 
let us consider again the fermionic particles in (\ref{particlegen}), 
which should span an SU(3) invariant subspace.
Hence, an SU(3) transformation acting on these states should be of the form,
\beqn
R_{I}\left(V^\pm_a\right)^\dagger \longmapsto \sum_J 
c_{IJ} R_{J}\left(V_a^\pm\right)^\dagger,
\eeqn
with $c_{IJ}\in\mathds{C}$. 
The transformation matrix itself will be a linear combination of the full set of basis elements (\ref{genbas}).
These basis elements act on the fermionic states as follows,
\beqn
R_{K}{R_J}^\dagger R_{I}\left(V^\pm_a\right)^\dagger = \delta_{IJ} R_{K}\left(V_a^\pm\right)^\dagger.
\eeqn
We can then deduce the form of SU(3) generators
by demanding that their action leaves the elements of (\ref{particlegen}) with $I=4$ invariant.
The most general expression satisfying this is,
\beqn
\label{lambdabar}
\bar{\lambda}_I=\sum_{K, L=1}^3 \Omega_{IKL} R_K (R_L)^\dagger\,, \qquad
\Omega_{IKL}\in \mathds{C}\,,\quad I=1,\hdots, 8\,.
\eeqn
The coefficients $\Omega_{IJK}$ are fixed by assigning the triplet charges to the index $I=1,2,3$ of our fermions. 
Then the $\bar{\lambda}_I$  become maps between the different colour charges, as desired.

The generators acting on the anti-particle states (\ref{antiparticlegen}) 
can then be derived by using (\ref{conjugategaugetransform}).
In order to ensure 
$(\mathcal{O}K)^{\bar{*}}=(e^{i\bar{\lambda}_I})^{\bar{*}}K^{\bar{*}}
 = e^{-i\bar{\lambda}_I^{\bar{*}}}K^{\bar{*}}\overset{!}{=}
\mathcal{O}K^{\bar{*}}$,
the anti-particle SU(3) generators must be given by $-\bar{\lambda}_I^{\bar{*}}$.
Now, due to (\ref{conjugateR}), $\bar{\lambda}_I$ and $\bar{\lambda}_I^{\bar{*}}$ are linearly independent. 
This seems problematic, as we wish to have only one set of $\mathfrak{su}(3)$ generators acting on 
both particle and anti-particle states. 
However, we note that $\bar{\lambda}_I$ and $\bar{\lambda}_I^{\bar{*}}$ commute and $\bar{\lambda}_I$ 
annihilates antiparticles while $\bar{\lambda}_I^{\bar{*}}$ annihilates particles under left multiplication.
Thus we may instead identify one set of generators given by,
\beqn
\lambda_I = \bar{\lambda}_I-\bar{\lambda}_I^{\bar{*}}\,,\qquad I=1,\hdots, 8\,,
\eeqn 
which satisfies (\ref{conjugategaugetransform}).
This finally gives precisely the expressions in (\ref{su3gen}). 
Now, $e^{i\lambda_I}$ correctly transform both particle 
and anti-particle states, and the generators $\lambda_I$ will be the ones 
associated to the gauge field of $\mathfrak{su}(3)$.

Note that the Higgs elements (\ref{particlehiggs}) and the corresponding complex conjugate elements 
are automatically invariant under these transformations, just like the singlet states of (\ref{sum:gen2}). 

The transformation properties of the particle states can be used to derive the transformation properties of the gauge generators themselves. The transformation properties and commutation relations of our generators are discussed in detail in appendix~\ref{app:bogg}\,.

\subsubsection{$\text{U}_Y(1)$ transformations}
Similarly to the SU(3) charges, hypercharges are also assigned to the index $I$
and thus the hypercharge transformation must also act form the left. 
Since the $\text{U}_Y(1)$ transformation does not transition between colours, 
it must be constructed only out of elements of the form $R_I(R_I)^\dagger$, which takes the index $I$ to the index $I$. 
Thus our hypercharge generator must be of the form
\beqn
Y=\sum_{I=1}^8 y_{I} R_I (R_I)^\dagger\,,\qquad y_I\in \mathds{C}\,,
\eeqn
where the $y_I$ will be fixed by the charge assignments.
We then arrive at the following generator on 
particle states (\ref{particlegen}) and (\ref{particlehiggs}),
\beqn
\bar{Y}=-R_4\left(R_4\right)^\dagger + \frac13\sum_{I=1}^3 R_{I}\left(R_I\right)^\dagger\,.
\eeqn
When acting on anti-particle states (\ref{antiparticlegen}) and (\ref{antiparticlehiggs}) from the left,
the hypercharge generator becomes $-\bar{Y}^{\bar{*}}$.
Just as with the $SU(3)$ generators, 
we then construct a combined generator which transforms both particle and antiparticle states 
and satisfies (\ref{conjugategaugetransform}),
\beqn
\label{truehyp}
Y =\bar{Y}- \bar{Y}^{\bar{*}}=R_8\left(R_8\right)^\dagger-R_4\left(R_4\right)^\dagger + \frac13\sum_{I=1}^3 R_{I}\left(R_I\right)^\dagger-\frac13\sum_{I=5}^7 R_{I}\left(R_I\right)^\dagger.
\eeqn
Note that the hypercharge generator defined in this way automatically commutes with the $SU(3)$ generators. 
This is crucial since both the SU(3) 
and $\text{U}_Y$(1) transformations are described by matrix multiplication from the left and we need to be 
able to treat them as independent.

\subsubsection{SU(2) transformations}
Finally, we turn our attention to the SU(2) transformations.
Since there are currently no chiral structures present, 
there is no notion of left and right-handed fermions. 
We will show that the particles and anti-particle states which transform under SU(3) and $\text{U}_Y(1)$  
can also be made to have consistent transformations under SU(2). To construct a SU(2) with the appropriate chiral discrimination one needs to include projectors from $\mathds{C}\otimes\mathds{H}$ in the $\mathfrak{su}$(2) generators such that SU(2) only acts on left chiral fermions. A similar procedure can be employed to yield chiral discrimination of U$_Y$(1) hypercharges. This procedure is outlined in appendix \ref{righthandedreps}. Since the focus of this present paper is only on gauge structures and not spatial representations, a full decomposition of spatial and gauge components is left for future work.

The operator of SU(2) transformations will
transform any particle state in (\ref{particlegen}) and (\ref{particlehiggs}) as,
\beqn
\label{su2transf}
R_{I}\left(V^\pm_a\right)^\dagger \longmapsto ~~c_+ R_{I}\left(V_a^{+}\right)^\dagger+c_{-}R_{I}\left(V_a^{-}\right)^\dagger,
\eeqn
with $c_\pm\in\mathds{C}$. 
Again we would like to express this operation in terms of matrix multiplication of the elements.
However, in this case we are forced to consider right multiplication, as the elements in 
(\ref{particlegen}) and (\ref{particlehiggs}) with $V_a^+$
and $V_a^-$ are in different left-invariant subspaces. 
Therefore, there is no matrix multiplication on the left which could transition between 
weak-isospin $\pm$ states.

When acting with $V_a^\pm\left(V_a^l\right)^\dagger $ on 
the particle states from the right, we have that,
\beqn
R_{I}\left(V^\pm_a\right)^\dagger V_b^\pm\left(V_c^l\right)^\dagger = \delta_{ab} R_{I}\left(V_c^{l}\right)^\dagger.
\eeqn
where we have used (\ref{orthoV}).
The SU(2) transformations should only affect the $\pm$ index and not the generation index $a$ on $V_a^\pm$.
Thus, the SU(2) generators must must not involve 
terms of the form $V_a^\pm V_b^\pm$ with $a\neq b$.
Assigning the index $+$ to states with weak-isospin up and $-$ to states with weak-isospin down then 
fixes the generators to be precisely of the form (\ref{su2genparticle}).

Using (\ref{conjugategaugetransform}) we find that when a particle state transforms as,
\beqn
M\longmapsto Me^{iT_j}\,,\qquad j\in\{1,2,3\}\,,
\eeqn
then the antiparticle state must transform as,
\beqn
M^{\bar{*}}\longmapsto M^{\bar{*}} e^{-iT_j^{\bar{*}}}\,,\qquad j\in\{1,2,3\}\,.
\eeqn
So far this is similar to what we saw for the SU(3) generators. 
However, while $\bar{\lambda}_I$ and $\bar{\lambda}_I^{\bar{*}}$ 
were linearly independent, the SU(2) generators satisfy,
\beqn
T_1^{\bar{*}}=T_1 \qquad T_2^{\bar{*}}=T_2 \qquad T_3^{\bar{*}}=-T_3.
\eeqn
Thus we cannot build linear combinations of generators as we did for SU(3) and $\text{U}_Y(1)$.
As a consequence, we must define the two separate matrix multiplications (\ref{su2transfpart}) and (\ref{su2transfantipart}) for SU(2) transformations. 
Interestingly, this means that SU(2) transformations of particle and 
anti-particle elements in M(8,$\mathds{C}$) have the same form as 
Lorentz rotations of Weyl spinors in $\mathds{C}\otimes\mathds{H}$, 
as detailed in appendix \ref{lorentztransfweyl}.

\subsection{Linear independence}
\label{appendix:linearindependence}

We now verify that all elements assigned to the 62 different particle types of the Standard Model can be made linearly independent. We emphasise again that as we are only looking at gauge structures appearing in $\mathds{C}\otimes\overleftarrow{\mathds{O}}\subset\overleftarrow{\mathds{D}}$ we will consider only complex subspaces and their linear independence, as distinguishing between real and complex parameters requires the inclusion of Lorentz representations in $\mathds{C}\otimes\mathds{H}\subset\overleftarrow{\mathds{D}}$.

The relevant subspaces are spanned by elements of the form:
\begin{subequations}
\begin{align}
\label{genandantigenbases}
\text{Generations \& anti-generations}: \qquad&\Big\{R_I\left(V^\pm_a\right)^\dagger\Big\}_{I=1}^{8}\quad\text{with }a=1,2,3
\\\label{higgs}
\text{Higgs doublet}: \qquad&\Big\{R_4\left(V^\pm_\phi\right)^\dagger\Big\}
\\\label{su3bases}
\text{SU(3) generators}:
\qquad&\Big\{\lambda_I\Big\}_{I=1}^8
\\\label{su2bases}
\text{SU(2) generators}:
\qquad&\Big\{T_j\Big\}_{j=1}^3
\\\label{hypbasis}
\text{Hypercharge generator}:
\qquad&Y
\end{align}
\end{subequations}
Note that by construction all generations and anti-generations are linearly independent from each other. 
We now use that
\begin{align}
\label{v0plus}
V_0^+ 
=\sum_{I=1}^8 a_I R_I\,,
\qquad
V_0^- = \sum_{I=1}^4 (a_{I+4})^* R_I + \sum_{I=5}^8 (a_{I-4})^* R_I\,,
\end{align}
where, for notational simplicity, we have renamed $a_{0I}^+\to a_I$. 
Our requirement (\ref{conjugateV}) fixes the coefficients in $V_0^-$.
Now, due to the orthogonality of the $V_a^\pm$ any element $K$ in (\ref{genandantigenbases}) satisfies
\beqn
K V_0^\pm = 0\,.
\eeqn
This observation provides us with a necessary and sufficient condition 
for any linear combination $S$ of elements in 
(\ref{higgs})-(\ref{hypbasis}) to be linearly independent both from each other and from (\ref{genandantigenbases}).
Namely we must have that at least one of $SV_0^+$ and $SV_0^-$ does not vanish.
In order to achieve this, we simply need to exclude those sets of $\{a_I\}$ for which there 
exists at least one linear combination $S$ such that $S V_0^\pm=0$.

To this end note that in the $R_I(R_J)^\dagger$ basis we may write 
any complex linear combination of $\mathfrak{su}(3)$ generators in block-diagonal matrix form as
\beqn
\lambda =\text{Diagonal}\left\{M,0,-M^\mathrm{T},0\right\} 
\eeqn
where the zeros are just numbers and $M$ is a general 3$\times$3 complex traceless matrix. 
We then write an arbitrary linear combination of (\ref{higgs})-(\ref{hypbasis}) as
\beqn
\label{lincomb}
S=\lambda + c_jT_j+ d^+R_4\left(V_\phi^+\right)^\dagger+d^-R_4\left(V_\phi^-\right)^\dagger+g Y,
\eeqn
with $c_i,d^\pm,g\in\mathds{C}$. 
We now need to find those $a_I$ 
for which $SV_0^\pm = 0$ if and only if $\lambda=0$ 
and $c_j=d^\pm=g=0$.

\subsubsection*{Ensuring $c_j=d^\pm=g=0$}
Let us define the two following vectors in $\mathds{C}^3$, 
\beqn
a:=\begin{pmatrix}
	a_1&a_2&a_3
\end{pmatrix}^\mathrm{T}\,,\qquad \bar{a}:= 
\begin{pmatrix}
a_5&a_6&a_7
\end{pmatrix}^\mathrm{T}\,.
\eeqn
The equation $SV_0^+ = 0$ is equivalent to two vector and two scalar equations,
\begin{subequations}
\begin{align}
\label{A1eq}
M a +\left(c_1+ic_2\right)\bar{a}^* +\left(c_3+\frac13g\right)a &= 0\,,
\\
\label{A2eq}
-M^\mathrm{T} \bar{a} +\left(c_1+ic_2\right)a^* +\left(c_3-\frac13g\right)\bar{a} &= 0\,,
\\
\label{A3eq}
\left(c_1+ic_2\right)a_8^* +\left(c_3-g\right)a_4+d^+\left(h_0^+\right)^* &= 0\,,
\\
\label{A4eq}
\left(c_1+ic_2\right)a_4^* +\left(c_3+g\right)a_8&= 0\,.
\end{align}
\end{subequations}
Similarly, $SV_0^-=0$ gives,
\begin{subequations}
\begin{align}
\label{B1eq}
M \bar{a}^* +\left(c_1-ic_2\right)a -\left(c_3-\frac13g\right)\bar{a}^* &= 0\,,
\\
\label{B2eq}
-M^\mathrm{T} a^* +\left(c_1-ic_2\right)\bar{a} -\left(c_3+\frac13g\right)a^* &= 0\,,
\\
\label{B3eq}
\left(c_1-ic_2\right)a_4 -\left(c_3+g\right)a_8^*+d^-h_0^+ &= 0\,,
\\
\label{B4eq}
\left(c_1-ic_2\right)a_8 -\left(c_3-g\right)a_4^*&= 0\,.
\end{align}
\end{subequations}
From equations (\ref{A1eq}) and (\ref{A2eq}) we find that,
\beqn
c_3 (a^\mathrm{T}\bar{a}) = -\frac12\left(c_1+ic_2\right)\left(|a|^2+|\bar{a}|^2\right)\,,
\eeqn
while equations (\ref{B1eq}) and (\ref{B2eq}) yield,
\beqn
c_3 (a^\dagger\bar{a}^*) = \frac12\left(c_1-ic_2\right)\left(|a|^2+|\bar{a}|^2\right)\,.
\eeqn
Together we then have that,
\beqn
\label{c3}
c_3 |a^\mathrm{T}\bar{a}|^2 = \frac12\left(|a|^2+|\bar{a}|^2\right)
\left(ic_1\,\text{Im}(a^\mathrm{T}\bar{a}) - i c_2 \,\text{Re}(a^\mathrm{T}\bar{a})\right)\,,
\eeqn
and
\beqn
\label{c1c2}
c_1\,\text{Re}(a^\mathrm{T}\bar{a})+ c_2 \,\text{Im}(a^\mathrm{T}\bar{a})=0\,.
\eeqn
Note that since $V_0^+$ and $V_0^-$ are orthogonal vectors, this implies that,
\beqn
a^\mathrm{T}\bar{a} = - a_4a_8\,.
\eeqn
Using (\ref{c3}) in $a_8$(\ref{A3eq}) + $a_4$(\ref{A4eq}) and $a_4^*$(\ref{B3eq}) + $a_8^*$(\ref{B4eq}), we solve for $c_1$ and $c_2$ as linear combinations of $d^+$ and $d^-$. Then using (\ref{c1c2}) with these solutions we find that,
\beqn
\label{dpdm1}
|a_8|^2 d^+\left(h_0^+\right)^*a_4^* = -|a_4|^2d^- h_0^+a_8\,.
\eeqn
On the other hand, taking the differences $a_8$(\ref{A3eq}) - $a_4$(\ref{A4eq}) and $a_4^*$(\ref{B3eq}) - $a_8^*$(\ref{B4eq}) and using only (\ref{c1c2}), we find that,
\beqn
\label{dpdm2}
|a_8|^2 d^+\left(h_0^+\right)^*a_4^* = |a_4|^2d^- h_0^+a_8\,.
\eeqn
Together equations (\ref{dpdm1}) and (\ref{dpdm2}) hold if and only if any of the following are true: 
$a_4a_8=0$, $h_0^+=0$, or $d^+ = d^- =0$. 
Clearly we must have $h_0^+\neq0$ for the Higgs doublet 
to be linearly independent from the (anti-)generations. 
Thus, in order for $d^+ = d^- =0$ to be the only solution to $SV^\pm_0=0$, we must demand that 
 $a^\dagger\bar{a}^*=-a_4^*a_8^*\neq0$. In other words, we find the conditions,
 \beqn
 \begin{pmatrix}
a_{1}&a_{2}&a_{3}
\end{pmatrix}^\mathrm{T}~&\not\perp~
\begin{pmatrix}
a_{5}^*&a_{6}^*&a_{7}^*
\end{pmatrix}^\mathrm{T}\,,
\qquad h_0^+\neq0\,.
 \eeqn
Since $c_1$ and $c_2$ are linear combinations of $d^+$ and $d^-$, 
they also both vanish. Then (\ref{c3}) implies that $c_3=0$, 
and equations (\ref{A3eq}), (\ref{A4eq}), (\ref{B3eq}), and (\ref{B4eq}) 
imply $g=0$, which hence does not need to be enforced separately.

\subsubsection*{Ensuring $\lambda=0$}

With $c_j=d^\pm=g=0$, the equations (\ref{A1eq})-(\ref{B4eq}) 
reduce to $\lambda V^\pm_0=0$. Writing any linear combination $\lambda$ in terms of its generators as
\beqn
\label{sumlambda}
\lambda =b_1\lambda_1 + b_2\lambda_2 + b_3\lambda_3 + b_5\lambda_5 + b_6\lambda_6 + b_7\lambda_7 + b_8\lambda_8\,,
\qquad b_i\in\mathds{C}\,,
\eeqn
we use the explicit form for $\lambda_I$ in (\ref{su3gen}) and define the two matrices,
\begin{align}
\label{mams}
m_\mathrm{S} \equiv
\begin{pmatrix}
b_3+b_8 &b_1 &b_4
\\
b_1 & b_8-b_3 & b_6
\\
b_4 & b_6 & -2b_8
\end{pmatrix}\,,
\qquad
m_\mathrm{A} \equiv
\begin{pmatrix}
0 &-b_2 & -b_5
\\
b_2 & 0 & -b_7
\\
b_5 & b_7 & 0
\end{pmatrix}\,.
\end{align}
It is straightforward to verify that the matrix equations $\lambda V_0^\pm=0$ are equivalent to,
\begin{align}
&a\,,\,
\bar{a}^*
~~\in~~ \text{Kern}\big(m_\mathrm{S}+im_\mathrm{A}\big) \cap \text{Kern}\big(m_\mathrm{S}-im_\mathrm{A}\big)\,.
\end{align}
Here, Kern($S$) denotes the kernel of $S$, and $\cap$ denotes the intersection of the two kernels.
This in turn implies,
\begin{align}
\label{kernelspace}
&a\,,\,
\bar{a}^*
~~\in~~ \text{Kern}\big(m_\mathrm{S}\big) \cap \text{Kern}\big(m_\mathrm{A}\big)\,.
\end{align}
It is easy to convince oneself that, 
since $m_S$ is a traceless symmetric matrix and $m_A$ is antisymmetric, neither of them can 
have rank 1.\footnote{If the symmetric matrix has has rank 1, 
it only has 1 non-vanishing eigenvalue and can therefore not be traceless.
For the antisymmetric matrix $m_A$, this can also be seen by noting that a rank-1 matrix can be written
in terms of 2 vectors $u$ and $v$ as $m_A=uv^\mathrm{T}$. 
But then antisymmetry implies $uu^\mathrm{T}=0$ and thus $u=0$.}
Hence, they must have rank 3, 2 or 0. 
The matrices in (\ref{mams}) cannot both be trivial, because this would imply $\lambda=0$.
If either of the matrices
has rank 3 then $\lambda V_0^\pm\neq0$ for any $V_0^\pm\neq0$
and there is nothing left to show. 
This leaves us with the case where at least one of the matrices has rank 2 and thus a kernel of dimension 1. 
Then for both vectors in (\ref{kernelspace}) to be in both kernels,
this implies that they must be proportional $a\propto \bar{a}^*$, as neither vector may be the zero vector for $c_j=d^\pm=g=0$.

Thus we find that the necessary and sufficient conditions for (\ref{genandantigenbases})-(\ref{hypbasis}) 
to be linearly independent are,
\beqn\label{propc}
\begin{pmatrix}
a_{1}&a_{2}&a_{3}
\end{pmatrix}^\mathrm{T}~&\not\propto&~
\begin{pmatrix}
a_{5}^*&a_{6}^*&a_{7}^*
\end{pmatrix}^\mathrm{T}\,,
\nonumber\\
\begin{pmatrix}
a_{1}&a_{2}&a_{3}
\end{pmatrix}^\mathrm{T}~&\not\perp&~
\begin{pmatrix}
a_{5}^*&a_{6}^*&a_{7}^*
\end{pmatrix}^\mathrm{T}\,,
\qquad
h^\pm_0\neq0\,.
\eeqn
Note that neither vector $a$ nor $\bar{a}$ may be the zero vector.
This concludes the analysis of linear independence and the derivation
of our main result.

\section{Discussion}
\label{discussion}

In this paper we have explicitly demonstrated that the complexified octonions, 
as the largest of the normed division algebras, have the potential to incorporate the entire Standard Model particle content
without requiring the introducing new gauge groups. 
Our result extends previous works, as already discussed in detail in section~\ref{previousresults}. 
To the best of our knowledge, we have presented here for the first time a direct sum decomposition of the algebra $\mathds{C}\otimes\overleftarrow{\mathds{O}}\subset\overleftarrow{\mathds{D}}$ which shows the gauge representations of the bosonic and fermionic particles of the Standard Model. We also comment on how one may include the algebra $\mathds{C}\otimes\mathds{H}$ to be able to simultaneously incorporate both left and right handed fermions with appropriate symmetry transformations.

We stress that our result is not a derivation of the Standard Model, 
but rather an identification of its particle content within the maps on the complexified octonions. 
Further work is needed to determine whether the Standard Model particle choice is unique within the algebra, 
 or whether additional assumptions need to be invoked in order to make the identification unique. 
 The automorphism group of the octonions is $\mathrm{G}_2$ and contains SU(3) as a subgroup,  
 implying that at least part of the Standard Model 
 gauge group is fundamental to the octonions themselves~\cite{Furey:2016ovx}. 
 This suggests that a derivation of the Standard Model particle 
 content from the octonions may indeed be possible. This is highlighted by the work \cite{Gunaydin:1973rs,Gunaydin:1974fb} that studies quark structures from the octonions. Additionally, recent developments, \cite{Todorov:2018yvi,Todorov:2019hlc}, use the exceptional Jordan algebras, formulated in terms of sets of octonions, as an approach to deriving both the Standard Model gauge groups and sets of generations. 
 
It could be enlightening to translate our result into the the Clifford
algebra formulation, using the isomorphism $\mathds{C}l(6)\cong\text{M(8,}\mathds{C})$. We discuss the basic 
ingredients of this formalism in appendix~\ref{app:cliffalg} and the isomorphism in \ref{appendix:matrixisomorphism}.
This would allow for a more direct comparison to the 
works in Ref.~\cite{Furey:2014iwa, Furey:2015tqa, Furey:2016ovx, Furey:2018yyy} 
and possibly make certain features
of our result more apparent. Nevertheless, we emphasize once more that one of the advantages of working with
the matrix algebra $\text{M(8,}\mathds{C})$ is its mathematical simplicity.

The decomposition of M(8,$\mathds{C}$) as per (\ref{decompM8}) highlights the 
existence of 2 elements which do not belong to the Standard Model. 
Consequently the additional components, lying in the space $P_\mathrm{BSM}$, are ideal candidates for new 
particles beyond the Standard Model. Since we only require the space $P_\mathrm{BSM}$ 
to be linearly independent from the Standard Model particles (and not orthogonal) 
it is currently not clear how to unequivocally assign gauge representations 
to these additional elements. Nevertheless, we expect them to be SU(3) singlets since 
there is no third element present to fill up another triplet representation.
It would be very interesting to further investigate the nature of the additional particles 
and determine whether there is a chance to observe them experimentally. 
A better understanding of their physical properties is also required in order to 
understand their possible effects on areas beyond particle physics, e.g.~cosmology.

We emphasize that including the spacetime structures described by $\mathds{C}\otimes\mathds{H}$ 
is necessary for a precise identification of the Standard Model including Lorentz representations. 
In order to address questions involving Lorentz structures (e.g. why left handed fermions transform as doublets of the weak SU(2) gauge group while right handed fermions are singlets), one has to study the full algebra $\overleftarrow{\mathds{D}}$~\cite{Furey:2016ovx}. 
Eventually, this may also offer more insight into the particle nature of $P_\mathrm{BSM}$ and we leave these
interesting investigations for future work.

When investigating the full Dixon algebra $\overleftarrow{\mathds{D}}$, there is a fundamental difference 
between the sub-algebras $\mathds{C}\otimes\mathds{H}$ and 
$\mathds{C}\otimes\overleftarrow{\mathds{O}}\cong\text{M(8,}\mathds{C})$.
 In $\text{M(8,}\mathds{C})$ we identified the Lie algebras generating the 
 Standard Model gauge group $G_\mathrm{SM}$. 
 We then constructed subspaces transforming in fundamental representations 
 of $G_\mathrm{SM}$. These subspaces are linearly independent from each other and from the Lie algebra. 
 The situation is different for Lorentz representations in $\mathds{C}\otimes\mathds{H}$. 
 As discussed in appendix~\ref{lorentztransfweyl}, 
 the Lie algebra $\mathfrak{so}(1,3)$, as a complex vector space, is spanned by 3 basis elements. 
 On the other hand, left- and right-handed Weyl spinors each span a complex 2-dimensional subspace of 
 $\mathds{C}\otimes\mathds{H}$. 
 Since $\mathds{C}\otimes\mathds{H}$ has only 4 complex dimensions, 
 the Lie algebra of Lorentz transformations and the space of Lorentz representations are not linearly independent.
This distinction may be crucial for a consistent formulation of the Standard Model within 
$\overleftarrow{\mathds{D}}$, incorporating both Lorentz and gauge structures.

We end the discussion by commenting on the obvious fact that our present 
approach does not include gravitational degrees of freedom. At this 
stage, it is not clear whether the Lorentz structures inside $\mathds{C}\otimes\mathds{H}$ 
require a field theory formulation in flat spacetime. In fact, it is 
also possible that they are related to the local Lorentz symmetry 
present in the tetrad formulation of General Relativity. These are 
interesting open question to pursue in the future.

\vspace{30pt}

\paragraph{Acknowledgements.} This work is supported by a grant from the Max-Planck-Society.

\newpage

\appendix
\section{The matrix algebra $\text{M(8,}\mathds{C})$}

\subsection{The Clifford algebra $\mathds{C}l(6)$}\label{app:cliffalg}
We briefly review basic properties of the 6-dimensional Clifford algebra.
All basis elements of the space $\mathds{C}l(6)$ can be generated by two sets of
vectors $\{\alpha_i\}_{i=1}^3$ and $\{\alpha_i^\dagger\}_{i=1}^3$,
where hermitian conjugation ${}^\dagger$ is an anti-automorphism,
implying $i^*=-i$ and $\alpha^* = -\alpha^\dagger$.
These generating vectors satisfy the anti-commutation relations,
\begin{align}\label{cralp}
\{\alpha_i,\alpha_j^\dagger\} = \delta_{ij}\,,\qquad 
\{\alpha_i,\alpha_j\}=0\,,\qquad 
\{\alpha_i^\dagger,\alpha_j^\dagger\}=0\,.
\end{align}
We will here not work with the elements of the generating space, 
but rather with the full set of basis elements which span the space $\mathds{C}l(6)$. 
To this end we define $\omega:=\alpha_1\alpha_2\alpha_3$ and the projectors
\begin{align}
P_0 := \omega^\dagger\omega\,,\qquad  
P_i:= \alpha_i\omega^\dagger\omega\alpha_i^\dagger\,.
\end{align}
Let moreover $\bar{P}_a := P_a^*$ for all $a\in\{0,1,2,3\}$.
It follows from (\ref{cralp}) that
$\omega$ is annihilated by right or left action of any $\alpha_i$. 
The projector $P_0$ was used in \cite{Furey:2016ovx} to find the Standard Model structure associated to one generation of fermions, but to our knowledge this is the first use of the projections $P_i$.

The above $8$ projectors are linearly independent and split our space into $8$ complex linearly independent subspaces. 
Specifically, each one of these projectors will define a left ideal. 
The space $\overleftarrow{\mathds{C}\otimes\mathds{O}}\,P_b$ can then be spanned by 8 linearly independent basis vectors,
\begin{align}
\label{basisvec1}
B_{ab} := \alpha_a\omega^\dagger\omega\alpha_b^\dagger;\qquad  A_{ab} := \alpha_a^\dagger\omega\alpha_b^\dagger\qquad\quad  a \in\{0,1,2,3\}\,.
\end{align}
The basis vectors which span $\overleftarrow{\mathds{C}\otimes\mathds{O}}P_b^*$ are found by taking the complex conjugate of (\ref{basisvec1}), which we will denote by
\begin{align}
\label{basisvec2}
\bar{B}_{ab} := \alpha_a^\dagger\omega\omega^\dagger\alpha_b\qquad \bar{A}_{ab} := \alpha_a\omega^\dagger\alpha_b\qquad\quad a \in\{0,1,2,3\}\,.
\end{align}
This provides a compact way of writing all basis elements of $\mathds{C}l(6)$ in terms of $B_{ab},\bar{B}_{ab},A_{ab},\bar{A}_{ab}$.

\subsection{$\mathds{C}l(6)\cong\text{M(8,}\mathds{C})$}
\label{appendix:matrixisomorphism}

We will now demonstrate that $\mathds{C}l(6)$ is isomorphic to the algebra of $8\times8$ complex matrices $\text{M}(8,\mathds{C})$. Clearly the vector spaces over which the two algebras are defined are isomorphic by virtue of having the same dimension. 
We thus only need to show that the Clifford product in $\mathds{C}l(6)$ is identified with the matrix product in 
$\text{M}(8,\mathds{C})$.

Let $\{M_{IJ}\}$ be a basis of $\text{M}(8,\mathds{C})$.
We take one of their matrix entries to be equal to 1 (in the $I$th row and $J$th column),
while all other entries are zero.
A general matrix $F$ in $\text{M}(8,\mathds{C})$ can then be written as $F=\sum_{I,J} F^{IJ}M_{IJ}$.
The matrix product expressed in this basis reads,
\beqn\label{matprod}
FH=\sum_{I,L}\left(\sum_JF^{IJ}H^{JL}\right)M_{IL}\,.
\eeqn
Next we identify the basis $M_{IJ}$ with the basis elements of $\mathds{C}l(6)$ via
\begin{align}
\label{matrixclifident}
M_{IJ} \longleftrightarrow\left\{
\begin{matrix}
B_{(I-1)(J-1)} & \text{for } I,J\in\{1,2,3,4\}
\\
A_{(I-5)(J-1)} &\qquad\qquad\quad\text{ }\text{ for } I\in\{5,6,7,8\}\text{, }J\in\{1,2,3,4\}
\\
\bar{A}_{(I-1)(J-5)} &\qquad\qquad\quad\text{ }\text{ for } I\in\{1,2,3,4\}\text{, }J\in\{5,6,7,8\}
\\
\bar{B}_{(I-5)(J-5)} &\text{for } I,J\in\{5,6,7,8\}
\end{matrix}
\right.
\end{align}
Under this identification, we can evaluate the Clifford algebra product of two basis elements and obtain,
\beqn
M_{IJ} M_{KL} = \delta_{JK} M_{IL}\,.
\eeqn
This reproduces precisely the standard matrix product of M(8,$\mathds{C}$) in (\ref{matprod}).

Hermitian conjugation of elements of $\mathds{C}l(6)$ correspond to the usual hermitian conjugation of matrices in $\text{M}(8,\mathds{C})$. Thus we will not distinguish between hermitian conjugation in the two algebras and label both the operations by ${}^\dagger$. Specifically, for any $M\in$ M(8,$\mathds{C}$) we have that
$M^\dagger := \left(M^*\right)^\mathrm{T}$ 
where ${}^\mathrm{T}$ is the matrix transpose.
The situation is different for the operation of complex conjugation. 
Due to the Clifford algebra property $\alpha^* = -\alpha^\dagger$, 
complex conjugation acts on the $\mathds{C}l(6)$ basis elements (\ref{basisvec1}) and (\ref{basisvec2}) as,
\begin{align}
\label{basisconjugation}
\left(B_{ab}\right)^* = \sum_{c,d} \eta_{ac}\bar{B}_{cd}\eta_{db}\,,\qquad
\left(A_{ab}\right)^* = \sum_{c,d} \eta_{ac}\bar{A}_{cd}\eta_{db}\,,
\end{align}
where $\eta$ is a diagonal matrix with entries $\{1,-1,-1,-1\}$.\footnote{Even though $\eta$ is the same as the Minkowski metric in Cartesian coordinates, this is just an artefact of how we chose to represent our basis elements, and not related to Lorentz transformations.}

From (\ref{matrixclifident}) and (\ref{basisconjugation}), 
it is clear that, in the matrix representation, complex conjugation necessarily 
affects the index structure. 
Namely, matrices $M\in\text{M}(8,\mathds{C})$ satisfy,
\begin{align}
\label{matrixrepconjugation}
M^{\bar{*}} :=
\begin{pmatrix}
0 &\eta
\\
\eta &0
\end{pmatrix}
M^*
\begin{pmatrix}
0 &\eta
\\
\eta &0
\end{pmatrix}\,,
\end{align}
where, in order to distinguish complex conjugation in the two algebras, 
we have introduced the symbol $\bar{*}$ to denote $\mathds{C}l(6)$ complex conjugation 
in the matrix representations.
We continue to use $^*$ to denote the conjugation of complex numbers.

As our matrix space can be written as the outer product of two vector spaces $\mathds{C}^{8}$, this implies that, for any $V\in\mathds{C}^8$, we have that,
\beqn\label{ccdef}
V^{\bar{*}} \equiv \begin{pmatrix}
	0 &\eta
	\\
	\eta &0
\end{pmatrix}
V^*.
\eeqn
In assigning basis elements of $\mathds{C}l(6)$ to Standard Model particle types we use the matrix representation $\text{M}(8,\mathds{C})$. This simplifies the analysis of linear independence and makes the paper more accessible to readers less familiar with Clifford algebras. However, we stress that we need properties, like the complex conjugation $\bar{*}$, associated to the complex Clifford algebra $\mathds{C}l(6)$.
The latter is isomorphic to the more fundamental structure $\mathds{C}\otimes\overleftarrow{\mathds{O}}$.

\section{Lorentz structures in $\mathds{C}\otimes\mathds{H}$}
\label{lorentztransfweyl}

We briefly outline in which way the complex quaternions
$\overleftarrow{\mathds{C}\otimes\mathds{H}}= \mathds{C}\otimes\mathds{H}$
contain the Lorentz representations of the Standard Model. 
For details, see Ref.~\cite{Furey:2016ovx}.
The basis elements of this algebra
are $\{1,i,\varepsilon_x,i\varepsilon_x,\varepsilon_y,i\varepsilon_y,\varepsilon_z,i\varepsilon_z\}$, 
where $\{\varepsilon_x,\varepsilon_y,\varepsilon_z\}$ anti-commute and satisfy
\begin{align}
\label{quatnormrules}
\varepsilon_x\varepsilon_y = \varepsilon_z\,, 
\qquad
\varepsilon_y\varepsilon_z=\varepsilon_x\,,
\qquad
\varepsilon_z\varepsilon_x=\varepsilon_y \,,
\qquad\varepsilon_i^2=-1\,.
\end{align}
The unit scalar, $1$, and the unit imaginary, $i$, commute with all other elements of the algebra. Hermitian conjugation ${}^\dagger$ can be defined on the algebra as an anti-automorphism which maps $\varepsilon_j\to-\varepsilon_j$ and $i\to-i$.

The above basis is useful for showing that $\mathds{C}\otimes\mathds{H}$ contains the Lie algebra $\mathfrak{so}(1,3)$ 
of Lorentz transformations. $\varepsilon_j$ generates rotations and $i\varepsilon_j$ generates boosts.
The same algebra contains Lorentz spinors and their transformations are described by letting the algebra act on itself.
To see this, it is convenient to work with the basis vectors,
\begin{align}
\nonumber
\varepsilon_{\uparrow\uparrow} := \frac12\left(1-i\varepsilon_z\right) \,,
\qquad \varepsilon_{\downarrow\downarrow} := \frac12\left(1+i\varepsilon_z\right)\,,
\\\label{quatspinorbasis}
\varepsilon_{\downarrow\uparrow} := \frac12\left(\varepsilon_y+i\varepsilon_x\right) \,,\qquad
\varepsilon_{\uparrow\downarrow} := \frac12\left(-\varepsilon_y+i\varepsilon_x\right)\,.
\end{align}
This basis satisfies,
\begin{align} 
\varepsilon_{\uparrow\uparrow}^2=\varepsilon_{\uparrow\uparrow}\,,\qquad
\varepsilon_{\downarrow\downarrow}^2=\varepsilon_{\downarrow\downarrow}\,,\qquad
\varepsilon_{\uparrow\downarrow}^2=\varepsilon_{\downarrow\uparrow}^2=0\,.
\end{align}
Defining $P := \epsilon_{\uparrow\uparrow}$,
left and right handed spinors correspond to the following subspaces,
\begin{align}
\Psi_L\in (\mathds{C}\otimes\mathds{H})P\,,
\qquad
\Psi_R\in (\mathds{C}\otimes\mathds{H})P^*\,.
\end{align}
Hence, $\epsilon_{\uparrow\uparrow}$ and $\epsilon_{\downarrow\downarrow}$ are projectors related by complex conjugation. 
Left and right handed Lorentz transformations of these spinors are defined as,
\begin{align}
\Psi_L \to e^{is}\Psi_L\,,
\qquad
\Psi_R \to e^{-is^*}\Psi_R\,,
\end{align}
for $s\in\mathfrak{so}(1,3)$ spanned by the elements $\varepsilon_j$ and $i\varepsilon_j$.
Then a general spinor behaves under Lorentz transformations as,
\begin{align}
\label{spinorlorentztransform}
\Psi \to e^{is}\Psi P + e^{-is^*}\Psi P^*.
\end{align}
Additionally within the algebra one can find scalar, vector, and field strength representations,  
as further elaborated on in Ref.~\cite{Furey:2016ovx}.

\section{Gauge transformations}
\label{app:bogg}

\subsection{Commuting transformations}

Here we discuss general conditions for gauge transformations acting on states 
such as (\ref{sum:gen}) and (\ref{sum:higgs}) to commute. 
Let us consider two gauge groups $G$ and $G'$ and denote their
gauge transformations by the operators $\mathcal{O}_G$ and $\mathcal{O}_{G'}$.
Then let some element $\Psi$ transform under these gauge groups as,
\beqn
\Psi\to \mathcal{O}_G\Psi\,,\qquad \Psi\to \mathcal{O}_{G'}\Psi\,.
\eeqn
If both of the operators are represented by matrix multiplication from the left, 
then,
\beqn
 \mathcal{O}_G\Psi =e^{i\lambda}\Psi\,,\qquad \mathcal{O}_{G'}\Psi=e^{i\lambda'}\Psi\,,
\eeqn
where $\lambda$ and $\lambda'$ are generators in the Lie algebras of $G$ and $G'$.
When these transformations arise from entirely independent gauge groups,
they should commute, 
\beqn
e^{i\lambda}e^{i\lambda'}\Psi \overset{!}{=} e^{i\lambda'}e^{i\lambda}\Psi\,.
\eeqn
This holds if and only if the generators $\lambda$ and $\lambda'$ commute. 
For instance, in our present setup, 
the generators of $\text{U}_Y(1)$ and SU(3) commute, and so do their gauge transformations
on all particle states.

On the other hand, 
suppose now that the transformations corresponding to the group $G$ are represented by matrix 
multiplication from the left and those of $G'$ 
are represented by matrix multiplication from the right,
\beqn
\mathcal{O}_G\Psi =e^{i\lambda}K\qquad \mathcal{O}_{G'}\Psi=\Psi e^{i\lambda'}.
\eeqn
In this case, as long as one works with an associative algebra, it does not matter which action is carried out first
since,
\beqn
e^{i\lambda}\Big(\Psi e^{i\lambda'}\Big)= \Big(e^{i\lambda}\Psi\Big)e^{i\lambda'}\,,
\eeqn
always holds true.
Hence, requiring gauge transformations to commute puts no constraints on the generators $\lambda$ and $\lambda'$.

\subsection{Transformations of generators}

Next let us see how gauge transformations act on generators that do not commute
with each other. For definiteness, let us take the gauge group to be $G=$\,SU($N$).
Let $\Psi$ be a state in the fundamental representation of $G$. 
A generator $\lambda\in\,\mathfrak{su}(N)$ then takes $\Psi$ to 
a different state $\tilde{\Psi}$ via the infinitesimal transformation, 
\beqn
\label{su3ab}
(1+i\lambda) \Psi = \tilde{\Psi}\,.
\eeqn
Consider then also $\Psi'$ and $\tilde{\Psi}'$ 
obtained via another, finite transformation generated by~$\lambda'$,
\beqn
\label{atoaprime}
\Psi' = e^{i\lambda'}\Psi\,,
\qquad 
\tilde{\Psi}' = e^{i\lambda'}\tilde{\Psi}\,.
\eeqn
We can now derive the generator $\tilde{\lambda}$ which infinitesimally 
relates the states $\Psi'$ and $\tilde{\Psi}'$. 
Combining (\ref{su3ab}) and (\ref{atoaprime}), one arrives at,
\beqn
\tilde{\Psi}'-\Psi' =i\tilde{\lambda}\Psi'=i\left(e^{i\lambda'}\lambda e^{-i\lambda'}\right)\Psi'  \,.
\eeqn
Of course this derivation is well-known and shows that elements $\lambda$ of the Lie algebra $\mathfrak{su}(N)$ 
transform in the adjoint representation of the gauge group.

We will use the same method to derive how the SU(2) generators $T_i$ in our setup (which do not commute with
the elements $\lambda_I$ of $\mathfrak{su}(3)$) transform under the SU(3) transformations. 
A particle state $\Psi$ 
is infinitesimally related to $\tilde{\Psi}$ 
with different SU(3) and SU(2) charges via, 
\beqn
\label{su2su3}
\Psi+i\lambda_I \Psi+ i\Psi T_i = \tilde{\Psi},
\eeqn
Consider then a finite SU(3) transformation generated by $\lambda'$
that produces the states $\Psi'=e^{i\lambda'}\Psi$ and $\tilde{\Psi}'=e^{i\lambda'}\tilde{\Psi}$, 
Following the same procedure as above, we find,
\beqn
\tilde{\Psi}'-\Psi'=i e^{i\lambda'}\lambda_Ie^{-i\lambda'}\Psi'+i\Psi'T_i \,.
\eeqn
This shows that, even though they do not commute with each other, 
the SU(2) generators do not transform under SU(3) transformations because they are not acting
on states from the same side.
It is easy to show that by the same argument the 
SU(3) generators and the $\text{U}_Y(1)$ generator are invariant under SU(2) transformations.  
We conclude that all gauge generators transform only under 
their own gauge group, in the adjoint representation, as expected.

\section{Right Handed Representations}
\label{righthandedreps}
We here show explicitly how one may accommodate right handed gauge representations in the set-up provided here. To do so we first start by discussing the discrimination of the SU(2) transformations on different matrix elements.

The SU(2) transformations presented in (\ref{su2transfpart}) and (\ref{su2transfantipart}) clearly act differently on matrix elements $R_I(V_a^\pm)^\dagger$ depending on whether $I\in\{1,2,3,4\}$ or $I\in\{5,6,7,8\}$ respectively. We could incorporate such a discrimination at the level of the generators by defining projectors that single out these relevant matrix subalgebras. To do so define simultaneous multiplication on M(8,$\mathds{C}$) from the left by $X$ and from the right by $Y$, for some $X,Y\in\text{M(8,}\mathds{C}$), as the operation $X|Y$. Explicitly, for any $K\in\text{M(8,}\mathds{C}$), $(X|Y)K:= XKY$. 
Defining the projectors
\beqn
\mathcal{R}:= \sum_{I=1}^{4} R_I (R_I)^\dagger
\\
\bar{\mathcal{R}}:= \mathcal{R}^{\bar{*}} = \sum_{I=5}^{8} R_I (R_I)^\dagger
\eeqn
we can then describe the SU(2) transformations via operators
\beqn
\hat{T}_j := \mathcal{R}|T_j-\bar{\mathcal{R}}|T_J^{\bar{*}}
\eeqn
such that (\ref{su2transfpart}) and (\ref{su2transfantipart}) can collectively be written as
\beqn
K\to e^{i\hat{T}_j}K
\eeqn
for any $K\in$(\ref{sum:gen}).%
\footnote{Note that this is also the case for the $\mathfrak{su}$(3) generators. However since SU(3) transformations act from the left and commute with the projectors $\mathcal{R}$ and $\bar{\mathcal{R}}$ this would imply that defining similar operators $\hat{\lambda}_I := \mathcal{R}\bar{\lambda}_I|1 - \bar{\mathcal{R}}\bar{\lambda}_I^{\bar{*}}|1 = \left(\bar{\lambda}_I-\bar{\lambda}^{\bar{*}}_I\right)|1=\lambda_I|1$, where we have used $\bar{\lambda}_I$ as defined in (\ref{lambdabar}).}

This idea can be extended further when we consider spatial representations by ensuring that only left handed particles and right handed anti-particles transform under SU(2). In this case, using the projectors $P$ and $\bar{P}$ introduced in appendix \ref{lorentztransfweyl}, we would write operators
\beqn
\hat{T}'_j := \left(\mathcal{R}|T_j\right)P-\left(\bar{\mathcal{R}}|T_J^{\bar{*}}\right)P^*,
\eeqn
and SU(2) transformations via operators
\beqn
e^{i\hat{T}'_j}.
\eeqn

For U$_Y$(1) transformations we may also describe the action of left vs. right handed fermions via projectors. Here the story is quite similar, but with a small twist. Note that for left handed fermions, the hypercharge is the average of the electric charges of the SU(2) doublet. As such defining a matrix element
\beqn
Q := -R_4(R_4)^\dagger -\frac13\sum_{I=1}^3 R_I(R_I)^\dagger-\frac23\sum_{I=5}^7 R_I(R_I)^\dagger,
\eeqn
it is clear that
\beqn
Y \equiv Q-Q^{\bar{*}}
\eeqn
Now, define projectors
\beqn
\mathcal{V} := \sum_a V^-_a(V^-_a)^\dagger
\\
\bar{\mathcal{V}} := \mathcal{V}^{\bar{*}} = \sum_a V^+_a(V^+_a)^\dagger,
\eeqn
in the same way we defined the projectors $\mathcal{R}$ and $\bar{\mathcal{R}}$. It is then straight forward to verify that the operator
\beqn
\hat{Y} := \left((Q-Q^{\bar{*}})|1\right)P + \left(Q|\mathcal{V}-Q^{\bar{*}}|\bar{\mathcal{V}}\right)P^*
\eeqn
yields the correct hypercharge assignments for both left and right handed fermions, all from application of $Q$ and its complex conjugate. This is similar to how we obtain the correct SU(2) transformations from application of $T_j$ and their complex conjugates.

While we here detail an approach for how the correct Standard Model charges may be incorporated for both left and right handed fermions, it is clear that the use of projectors as introduced here is ad-hoc and not natural. However, it is the purpose of this paper only to show that the structure of $\mathds{C}\otimes\overleftarrow{\mathds{O}}$ when used as a subset of $\overleftarrow{\mathds{D}}$ may describe all Standard Model gauge representations, as presented here. Studies of the full $\overleftarrow{\mathds{D}}$ where the Lorentz structures are included explicitly, and thus any questions regarding the natural appearance of the Standard Model structures, is left to future work.


\end{document}